\documentclass{elsart}

\usepackage{amssymb,epsfig}

\begin{document}

\begin{frontmatter}



\title{Microscopic Magnetic Properties of
(V$_{1-x}$Ti$_{x}$)$_2$O$_3$ near the Phase Boundary of\\
the Metal-Insulator Transition}


\author{Jun Kikuchi\corauthref{cor1}},
\ead{kikuchi@ph.noda.sut.ac.jp}
\author{Noriyuki Wada},
\author{Kousuke Nara},
\author{Kiyoichiro Motoya}
\corauth[cor1]{Corresponding author (FAX: +81-471-23-9361)}

\address{Department of Physics, Faculty of Science and Technology,\\
Science University of Tokyo, Yamazaki, Noda, Chiba 278-8510,
Japan}

\begin{abstract}

Magnetic susceptibility ($\chi$) and $^{51}$V NMR have been measured in
(V$_{1-x}$Ti$_{x}$)$_2$O$_3$ near the phase boundary of the
metal-insulator transition.  It is established that the transition
from antiferromagnetic insulating (AFI) to antiferromagnetic metallic
phases near $x_{\rm c}\approx 0.05$ is not quantum critical but
is discontinuous with a jump of the transition temperature.  In the AFI
phase at 4.2 K, we observed the satellite in the zero-field $^{51}$V
NMR spectrum around 181 MHz in addition to the ``host'' resonance
around 203 MHz.  The satellite is also observable in the paramagnetic
metallic phase of the $x$ = 0.055 sample.  We associated the satellite
with the V sites near Ti which are in the V$^{3+}$-like oxidation
state but has different temperature dependence of the NMR shift from
that of the host V site.  The host $d$-spin susceptibility for $x$ =
0.055 decreases below $\sim$60 K but remains finite in the
low-temperature limit.

\end{abstract}

\end{frontmatter}

\section{Introduction}
\label{sec:intro}
Magnetic excitations from the anomalous metallic phase
of V$_2$O$_3$ and its derivatives have received continued interest for
more than three decades in connection with a famous
temperature-induced metal-insulator transition (MIT) \cite{mcwhan73}. 
Among various methods to stabilize the metallic phase against the
antiferromagnetic insulating (AFI) phase, introducing V vacancies
has extensively been studied in the last decade
\cite{carter91,bao93,yueda95,langenbuch96}.  While stoichiometric
V$_2$O$_3$ remains paramagnetic once the metallic phase is stabilized
by applying pressure \cite{gossard70,takigawa96,metoki01}, the ground
state of metal-deficient V$_{2-y}$O$_3$ is a spin-density wave (SDW)
state characterized by a small ordered moment ($\sim$0.15 $\mu_{\rm
B}$) with a large fluctuating component ($\sim$0.32 $\mu_{\rm B}$)
\cite{bao93}.  $^{51}$V NMR, on the other hand, has revealed the
existence of V$^{4+}$-like magnetic ``guest'' sites in V$_{2-y}$O$_3$
\cite{yueda78,langenbuch96}, indicating a spatially inhomogeneous
electronic state which is suspected to be incompatible with
homogeneous SDW order \cite{metoki01}.

Substitution of Ti for V also stabilizes the metallic phase,
resulting in the SDW ground state similar to V$_{2-y}$O$_3$
\cite{motoya98}.  However, magnetic guest sites have not been detected yet
in Ti-doped V$_2$O$_3$, which makes it a puzzle whether the SDW order is
intrinsic to a homogeneous system or to the system with some sort of
inhomogeneities like magnetic guests.  In this paper, we report on the
magnetic susceptibility and $^{51}$V NMR measurements on
(V$_{1-x}$Ti$_{x}$)$_2$O$_3$ near the MIT phase boundary.  We found
the V$^{3+}$-like guest sites distinguishable from the host site in
both the AFI and paramagnetic metallic (PM) phases.  The SDW order is
therefore intrinsic to the {\it doped} V$_2$O$_3$ systems with a kind of
magnetic impurities.

\section{Experiments}
\label{sec:expt}
Powder samples of (V$_{1-x}$Ti$_{x}$)$_2$O$_3$ were prepared by
arc-melting of V$_2$O$_3$, Ti and TiO$_2$ followed by annealing in
H$_2$ atmosphere at 950 $^{\circ}$C for 48 hours.  Magnetic
susceptibility was measured using a SQUID magnetometer.  A clamp-type
Cu-Be cell suited for the SQUID magnetometer \cite{uwatoko98} was used
for the measurement of the susceptibility under pressure.  NMR
measurements were carried out with a coherent-type pulsed
spectrometer.

\section{Results}
\label{results}
\subsection{Magnetic susceptibility and phase diagram}
\begin{figure}
\begin{center}
\epsfxsize=95mm \epsfbox{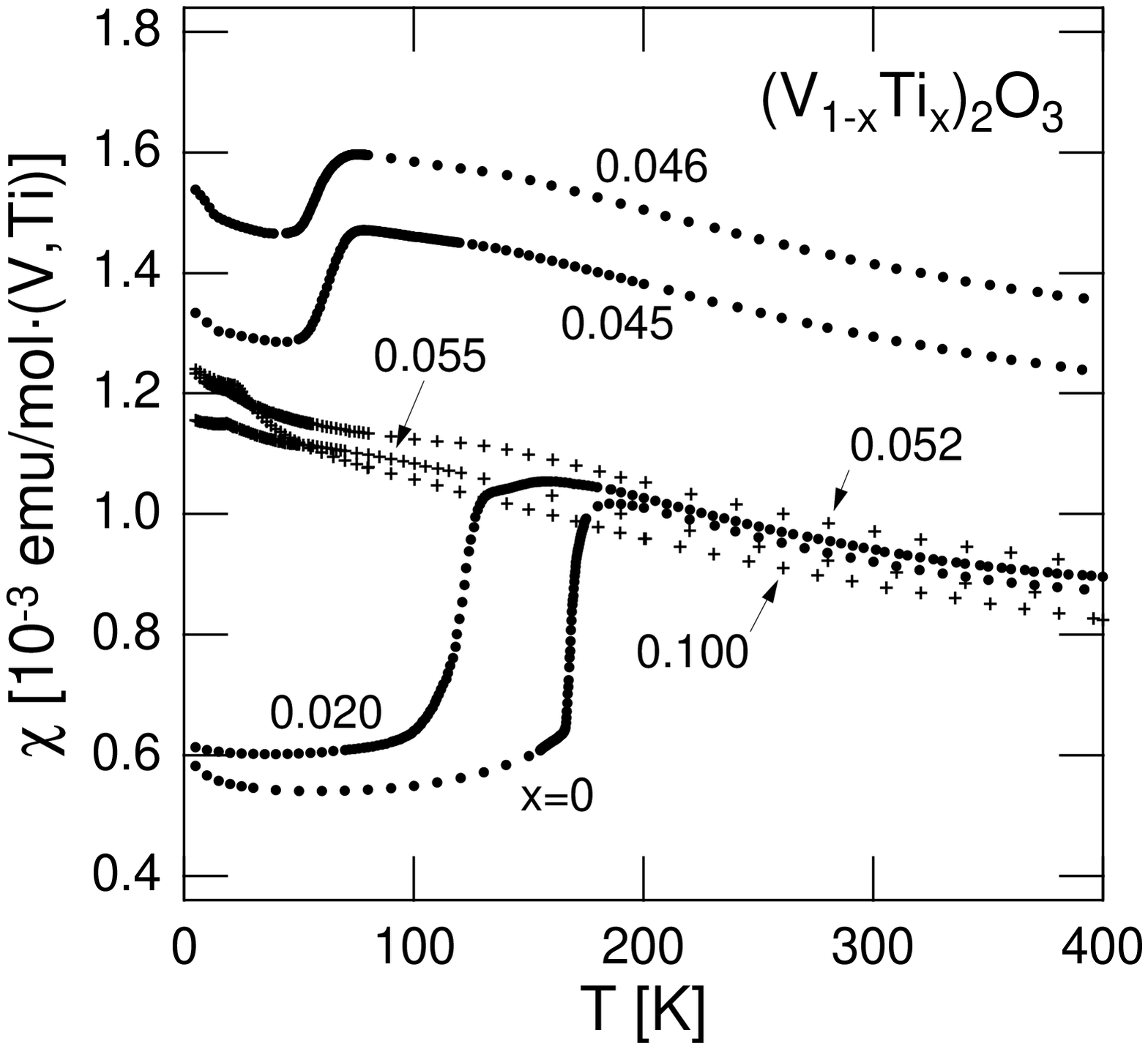}
\end{center}
\caption{Magnetic susceptibility of (V$_{1-x}$Ti$_{x}$)$_2$O$_3$.}
\label{fig:suscept}
\end{figure}

\vspace{-16pt}
Figure\ \ref{fig:suscept} shows the temperature ($T$) dependence of the magnetic
susceptibility $\chi$ = $M/H$ ($M$ and $H$ being the magnetization and
the external field, respectively) of (V$_{1-x}$Ti$_{x}$)$_2$O$_3$
measured at $H = 2$ T on heating. The results are in good agreement with those
reported in the literature \cite{yueda80} except that $\chi$ for $x$ =
0.045 and 0.046 are somewhat larger than the others.  This comes from
a nonlinear increase of $M$-$H$ curves at low fields for these samples due
to a trace of ferromagnetic impurities, of which existence does not
affect seriously the determination of transition temperatures because
their response to the external field is scarcely dependent on $T$.

\begin{figure}
\begin{center}
\epsfxsize=85mm \epsfbox{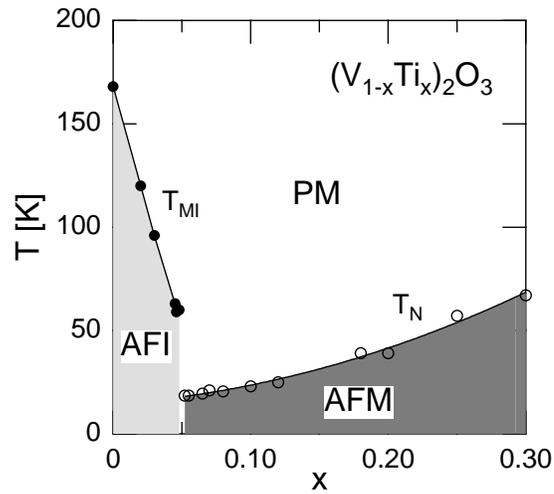}
\end{center}
\caption{Magnetic phase diagram of (V$_{1-x}$Ti$_{x}$)$_2$O$_3$.  AFM
denotes the antiferromagnetic metallic phase.}
\label{fig:diagram}
\end{figure}

We can see from Fig.\ \ref{fig:suscept} that a sudden drop of $\chi$
signaling MIT disappears for $x\geq 0.052$.  A kink then appears at low $T$
due to SDW ordering. The $T$-$x$ phase diagram is shown in Fig.\
\ref{fig:diagram}.  Here we defined the MIT temperature $T_{\rm MI}$
as a mid-point temperature of the rapid drop of $\chi$ due to MIT.
$T_{\rm MI}$ decreases with $x$ almost linearly to $\sim$60 K in a rate
$dT_{\rm MI}/dx \approx -23$ K/Ti\,\% and is terminated at the
critical concentration $x_{\rm c}\approx 0.05$.  Above $x_{\rm c}$,
N\'{e}el temperature $T_{\rm N}$ appears at $\sim$18 K and increases
gradually with $x$.  The phase diagram clearly demonstrates that the MIT
in (V$_{1-x}$Ti$_{x}$)$_2$O$_3$ with changing $x$ is not quantum
critical but is strongly first order.

\begin{figure}
\begin{center}
\epsfxsize=95mm \epsfbox{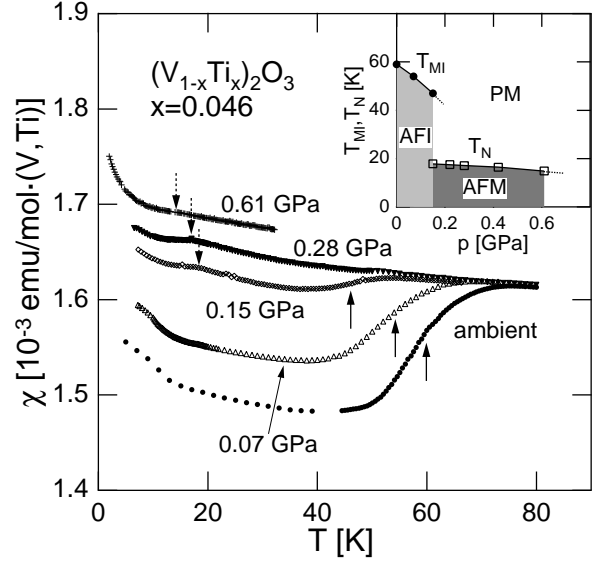}
\end{center}
\caption{Pressure dependence of the magnetic susceptibility of
(V$_{0.954}$Ti$_{0.046}$)$_2$O$_3$.  $T_{\rm MI}$ and $T_{\rm N}$ are
shown respectively by the solid and dashed arrows.  The data
are properly offset for clarity.  Inset: magnetic phase diagram under
pressure.}
\label{fig:press}
\end{figure}

The bulk $\chi$ of (V$_{0.954}$Ti$_{0.046}$)$_2$O$_3$ under pressure 
is shown in Fig.\ \ref{fig:press}.  The most striking feature of the
result is that both the anomalies due to MIT and SDW ordering are
observable at 0.15 GPa.  This suggests that the insulating and metallic
phases coexist in some pressure region, i.e., MIT under pressure is
first order as Ti doping.  The resulting $T$-$p$ phase diagram is
shown in the inset of Fig.\ \ref{fig:press}.  Applying pressure
reduces $T_{\rm MI}$ in a rate $dT_{\rm MI}/dp \approx -80$ K/GPa
which yields a scaling 0.29 GPa $\approx$ Ti\,\% for the MIT. As
opposed to Ti doping, $T_{\rm N}$ is suppressed by pressure, but the
rate $dT_{\rm N}/dp = -5.3(2)$ K/GPa is only one-fifth of that in
V$_{2-y}$O$_3$ \cite{carter91}.  $dT_{\rm N}/dp$ was found to decrease
slightly with $x$ to $-4.6(2)$ K/GPa for $x=0.065$.

\subsection{$^{51}$V NMR in the paramagnetic metallic phase}
\begin{figure}
\begin{center}
\epsfxsize=85mm \epsfbox{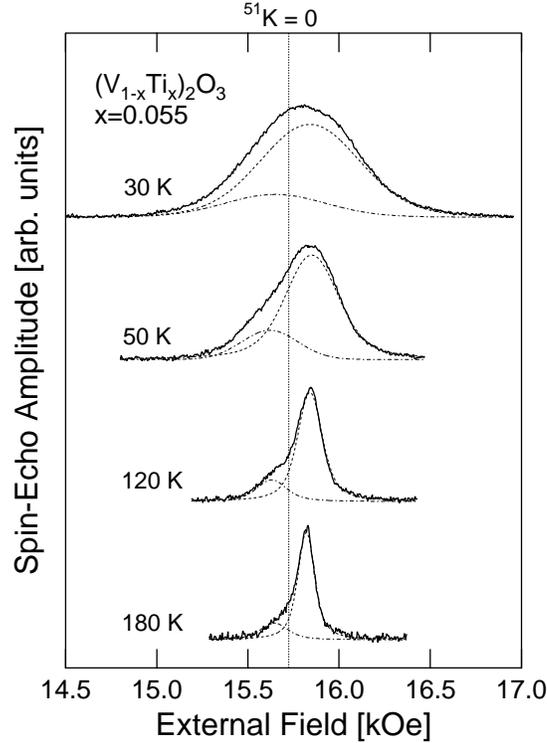}
\end{center}
\caption{Field-swept $^{51}$V NMR spectra in the paramagnetic state of
(V$_{0.945}$Ti$_{0.055}$)$_2$O$_3$ taken at 17.6 MHz.  The vertical
line indicates the field for zero shift.  The dashed and dashed-dotted
lines are the result of deconvolution (see text).}
\label{fig:spectraPM}
\end{figure}

\vspace{-16pt}
Field-swept $^{51}$V NMR spectra in the PM phase of
(V$_{0.945}$Ti$_{0.055}$)$_2$O$_3$ are shown in
Fig.\ \ref{fig:spectraPM}.  The spectrum is asymmetric with a tail at
lower fields and is broadened significantly at low $T$.  The
asymmetric line profile suggests the existence of V sites having different
crystallographic and/or magnetic environment from that in pure
V$_2$O$_3$.  We analyzed the spectrum by fitting to the sum of two
Voigt functions on the assumption that there exists another V site or
the ``guest'' site in addition to the ``host'' site.  The results are
also shown in Fig.\ \ref{fig:spectraPM}.  The intensity ratio between
the host (dashed line) and the satellite (dashed-dotted line
corresponding to the guest) resonances is about $4:1$ and is almost
$T$-independent.  This ratio can be understood by considering that the
satellite signal at lower fields comes from V nuclei with at least one
Ti atom on the four neighboring V sites, because in that case the
intensity ratio is $0.797:0.203 = 3.93:1$, agreeing well with the
observed one.  The guest site should be in the V$^{3+}$-like
electronic state as the host site, because Ti atoms exist as trivalent
ions in (V$_{1-x}$Ti$_{x}$)$_2$O$_3$ \cite{mcwhan73}.

\begin{figure}
\begin{center}
\epsfxsize=100mm \epsfbox{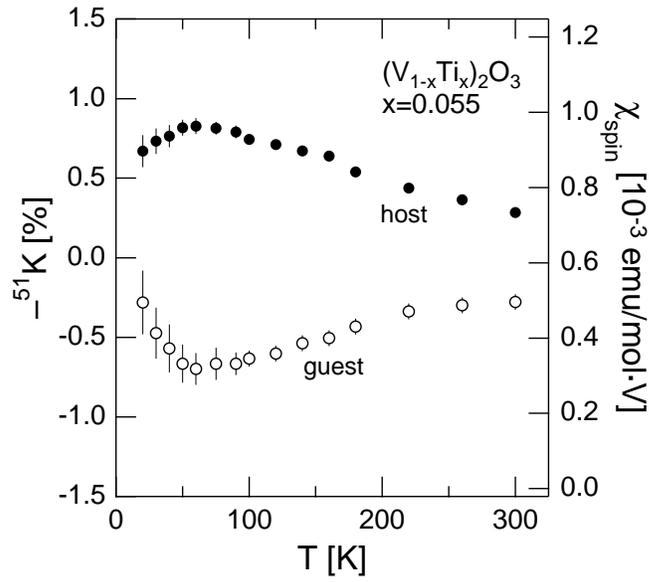}
\end{center}
\caption{Temperature dependence of the NMR shifts at the host ($\bullet$)
and guest ($\circ$) V sites in (V$_{0.945}$Ti$_{0.055}$)$_2$O$_3$.}
\label{fig:shift}
\end{figure}

The $T$ dependence of the NMR shifts $K^{h}$ and $K^{g}$  ($h$ and $g$
denote host and guest, respectively) at the V sites is shown in Fig.\
\ref{fig:shift}.  $-K^{h}$ takes a maximum around 60 K, following the NMR
shift of $^{51}$V in V$_2$O$_3$ under pressure \cite{takigawa96} as in
V$_{2-y}$O$_3$ \cite{yueda95}.  On the other hand, $-K^{g}$ decrease
gradually with decreasing $T$ and exhibits an upturn below about 60 K.
The different $T$ dependence of $K^{h}$ and $K^{g}$ indicates that the
$d$-spin susceptibilities of host and guest sites behave rather differently.
We also showed in Fig.\ \ref{fig:shift} the scale of the $d$-spin
susceptibility $\chi_{\rm spin}$ for the observed shift $K$ calculated
using the relation $\chi_{\rm spin} = (K-K_{\rm orb})/A_{\rm spin}$
($i=h,g$).  Here $K_{\rm orb}$ is the shift due to Van-Vleck orbital
susceptibility and $A_{\rm spin}$ is the $d$-spin hyperfine coupling
constant, for which the values in V$_2$O$_3$, $A_{\rm spin} = -132$
kOe/$\mu_{\rm B}$ and $K_{\rm orb} = 1.45$ \% \cite{takigawa96} were
used.  Noted that the host $\chi_{\rm spin}$ never vanishes in the
$T\rightarrow 0$ limit after making a round maximum around 60 K as in
V$_2$O$_3$ under pressure \cite{takigawa96}.  The origin of the $T$
dependence of $\chi_{\rm spin}$ for the host as well as the guest
sites is not clear at present and remains as a future problem.

\subsection{$^{51}$V NMR in the antiferromagnetic insulating phase}
\begin{figure}
\begin{center}
\epsfxsize=85mm \epsfbox{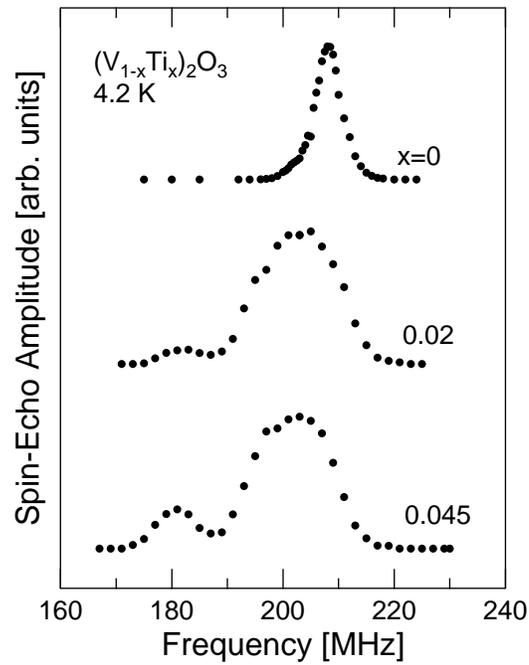}
\end{center}
\caption{Zero-field $^{51}$V NMR spectra in the antiferromagnetic insulating
phase of (V$_{1-x}$Ti$_{x}$)$_2$O$_3$ at 4.2 K.}
\label{fig:spectraAFI}
\end{figure}

\vspace{-16pt}
We have also observed the satellite in the zero-field $^{51}$V NMR spectrum in
the AFI state.  As shown in Fig.\ \ref{fig:spectraAFI}, the satellite
appears at lower frequencies of the host resonance.  The intensity of
the satellite increases with $x$, suggesting that the satellite signal
comes from V nuclei near Ti atoms.  The hyperfine fields determined
from the peak positions of the spectrum are listed in Table\
\ref{tab:hyper}.

A periodic modulation of the spin-echo amplitude when varying the rf-pulse
separation was observed for both the host and satellite resonances. 
From the modulation period, one can determine the splitting of the
quadrupolar septet of $^{51}$V ($I$=7/2) \cite{abe66}.  For $x$ = 0.02
the quadrupolar splitting of the host resonance is 218(2) kHz which is
slightly larger than that of pure V$_2$O$_3$ (200 kHz)
\cite{yasuoka71}.  A smaller splitting of 135(5) kHz
was found for the satellite in the $x$ = 0.02 sample.

The appearance of the satellite near the host signal is indicative of the
V$^{3+}$-like electronic state of the guest site in the AFI phase as in
the PM phase.  This is contrasted with the observation in
V$_{2-y}$O$_3$ where the V$^{4+}$-like guests are identified via the
zero-field $^{51}$V NMR \cite{yueda78,langenbuch96}.  We should note
here that the relative intensity of the satellite to the host signal
is consistent with the probability of finding V nuclei having Ti atoms
on the four neighboring V sites as in the PM phase.  It is therefore
reasonable to consider that the existence of V$^{3+}$-like guest sites is
characteristic of (V$_{1-x}$Ti$_{x}$)$_2$O$_3$ throughout its magnetic
phase diagram.
\begin{table}[t]
\caption{Hyperfine fields at the V sites in
(V$_{1-x}$Ti$_{x}$)$_2$O$_3$ in the antiferromagnetic insulating state
at 4.2 K.} \vspace{9pt}
\begin{tabular}{ccc} \hline
&\multicolumn{2}{c}{hyperfine field (kOe)}\\
Ti (\%) & host & satellite \\ \hline
0  & 185.9	& -\\
2	& 181.3	& 162.6\\
3	& 181.4	& 162.3\\
4.5	& 180.4	& 161.6\\
4.8	& 180.0	& 161.4\\ \hline
\end{tabular}
\label{tab:hyper}
\end{table}

\section{Discussion and Summary}
\label{discuss}
The observation of the V$^{3+}$-like guest sites demonstrates that even in
the metallic phase, the electronic state of (V$_{1-x}$Ti$_{x}$)$_2$O$_3$
is not so homogeneous as has been previoulsy thought.  In addition, the
host site preserves the V$_2$O$_3$ character in spite of the fact that
the number of $d$ electrons per vanadium is formally reduced by
doping.  These are nothing but a manifestation of strong Coulomb
repulsion between electrons which favors the number of electrons per
site being an integer.  The Coulomb interactions as well as nesting of
the Fermi surface are therefore indispensable for the SDW ordering in
metallic V$_2$O$_3$.

It seems, however, that the presence of the magnetic guest sites is
crucial in realizing the SDW order, because long-range order (LRO) is
absent in pure V$_2$O$_3$ despite the nearly identical spin
fluctuations in the metallic phase of pure, V-deficient and Ti-doped
V$_2$O$_3$ \cite{bao97,motoya98}.  A possible scenario is that
metallic V$_2$O$_3$ is unstable to a perturbation destroying spatial
coherence such as defects and impurities owing to the quantum critical
nature of spin fluctuations \cite{bao97}.  This is reminiscent of
impurity-induced LRO in low-dimensional quantum antiferromagnets with
disordered ground states \cite{renard95,azuma97}.  In metallic V$_2$O$_3$,
V-vacancy induced V$^{4+}$-like sites and Ti-doping induced
V$^{3+}$-like sites may serve as magnetic impurities, transforming the
system into the SDW state.  Local change due to doping of orbital
fluctuations suppressing spin order \cite{rice95} is also a likely
source for long-range SDW ordering in the doped V$_2$O$_3$ systems.

In summary, we established the first-order doping-induced and
pressure-induced MIT from AFI to AFM phases in
(V$_{1-x}$Ti$_{x}$)$_2$O$_3$.  The V$^{3+}$-like magnetic guest sites
exist in the PM phase as well as in the AFI phase, indicating the importance
of spatially non-uniform electronic states for the SDW ordering.

\section*{Acknowledgement}
\label{acknowledge}
One of the authors (J.K.) was supported by Grant-in-Aid for Encouragement of
Young Scientists from the Japan Society for the Promotion of Science
(No.~12740218).



\end{document}